email:jugao@uiuc.edu

# Classical Field Theory of Cerenkov Radiation Devices

Fang Shen and Ju Gao

*Department of Electrical and Computer Engineering,*

*University of Illinois, Urbana, IL 61801*

(Dated: January 25, 2006)

## Abstract

The usually continuous and incoherent Cerenkov radiation (CR) is converted into coherent and narrow-band radiation from within a dielectric-line cavity excited by a dc electron beam. We have studied the self-oscillating behavior of a device that operates at $\sim 22$ GHz by calculating explicitly the electron self-bunching and radiation growing processes. The nonlinear gain of the device and the radiation spectrum are also calculated. The agreement to the experiments indicates that the developed formalism could be used in designing a THz source based on cavity Cerenkov radiation.

The generation of radiation fields across the spectrum impacts almost all disciplines of science and technology. After masers and lasers, the next challenging frequency territory lies in the range between a few hundred GHz to THz. This is an important band for chemistry, biology, medicine and others because it matches the spectral range of the vibrational and rotational structures for vast amount of molecules. The development paths ultimately fall into two: one uses electrons in bound state for radiation and the other uses free electrons. The two paths reflect the basic radiation principle described by the Maxwell equations:

$$\nabla \times \mathbf{H} - \epsilon_0 \frac{\partial \mathbf{E}}{\partial t} = \mathbf{J} + \frac{\partial \mathbf{P}}{\partial t}, \tag{1}$$

where the magnetic and electric fields $\mathbf{H}$ and $\mathbf{E}$ are generated by either current density $\mathbf{J}$, which is made up of free electrons, or electric polarization $\mathbf{P} = \sum_i \mathbf{P_i}$ that is the sum of bound electrons.

Though devices based on bound electron radiation such as the cascade quantum well semiconductor lasers [1] are promising, devices based on free electrons possess some intrinsic advantages. When travelling in vacuum, these devices are exempt from thermal issues in materials and can therefore be operated in room temperature. They also offer broad bandwidth with easy tunability. A fact to note though is that the free-travelling electron does not radiate in vacuum, unless the radiation field is slowed down from the speed of light $c$. Various slow wave structures are thus invented, but one of the most straightforward mechanisms is using dielectric materials. Radiation occurs when the electron travels inside or in the vicinity the dielectric with the speed $u$ greater than the radiation field phase velocity $v_p$. The effect is known as the Cerenkov radiation (CR), illustrated by Fig. 1. Since its discovery [2, 3], CR has played an important role in high-energy physics, but the broad spectrum ranging from microwave to ultraviolet has also stimulated thoughts [4] of using CR for radiation generation, particularly in a frequency range that is difficult to access by other means.

CR is typically continuous and incoherent. Narrow-band and coherent radiation sources are useful and necessary. Such CR are generated by using prebunched electron beams [5] up to 30 GHz, but an external oscillation source is needed, e.g., a klystron, to define the spatial oscillation of the bunched beam. It is later demonstrated that coherent and narrow-band high frequency microwave radiation [6] can be generated by injecting an unbunched (dc) electron beam into a vacuum tunnel inside a dielectric-lined cavity as shown by Fig. 1(b).

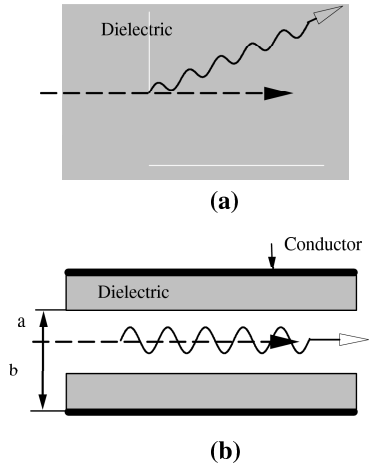

FIG. 1: Cerenkov radiation in (a) a dielectric medium and (b) a dielectric-lined cavity. The dashed lines represent electrons and the wiggled lines represent generated photons.

The device thus self-oscillates and owns the names of Cerenkov maser or laser.

Oscillation frequencies over 20 GHz have been achieved for the fundamental mode and over 150 GHz have been measured for the higher order modes. Based on waveguide theories, even higher frequency radiation such as THz radiation can be achieved by reducing the size of the device. The dielectric based structure makes it particularly interesting for the frequency scaling-up because the high precision micro-machining technologies are readily available nowadays.

That the CR device oscillates is interesting because the device does not possess an embedded "master oscillator" such as atoms in a laser [7] or, for that matter, alternating magnets in a free-electron laser [8]. The basic understanding about the mechanism has been reached in the literature [9–14] that the electrons generate the field within the cavity which in turn bunches the electron beam through interaction. Coherent radiation is then generated as a result of the self-bunched electron beam. The fundamental approach [15, 16] by a classical theory is using the two sets of equations: the Newton-Lorentz equation to calculate the electron motion with a given field and the Maxwell equations to solve for the field with the given electron motion. The two sets of the equations have to be solved simultaneously together with the correct boundary conditions. A device similar to that shown in Fig. 1 (b) is analyzed that operates at 8.75 GHz. In that model [15], the vacuum tunnel for the electron beam is omitted, which not only changes the boundary conditions but also is an indispensable feature for a practical device of eliminating electrons scattering off the materials. It is conceivable that the formulae developed there can be modified to analyze this device, we have chosen to develop even simpler equations with some different concepts that is perhaps more physically insightful.

We start with the single electron behavior inside the field, which is governed by the Newton-Lorentz equation:

$$
\begin{aligned}
d(\gamma m \mathbf{u})/dt &= -e\mathbf{E}(\omega, k) \\
d(\gamma mc^2)/dt &= -e\mathbf{u} \cdot \mathbf{E}(\omega, k),
\end{aligned}
\tag{2}
$$

where $e$ is the electronic charge. $\gamma m \mathbf{u}$ and $\gamma mc^2$ are the electron momentum and energy, respectively, and $\gamma = \frac{1}{\sqrt{1-u^2/c^2}}$. The field $\mathbf{E}(\omega, k)$ is the allowed mode inside the cavity. Here TM modes are considered because they dominate the interaction because $\mathbf{E} \parallel \mathbf{u}$. We choose to express the fields with real domains, avoiding using phasors,

$$
\mathbf{E}(\omega, k) = \mathbf{z} A(z) E(r) \cos(\omega t - kz),
\tag{3}
$$

where $\omega$ and $k$ are the frequency and wave constant of the field, respectively, whose relation, the dispersion relation, is to be determined by the cavity. Here $k$ is real, thus the growth or decay of the field is strictly described by the amplitude function $A(z)$ which takes no particular form and is to be solved by the coupled Newton-Lorentz and Maxwell equations. This treatment allows us to discuss the nonlinear behavior of the device. $E(r)$ is the radial function of the electric field that is solved by the current-free Maxwell equations.

The change of the electron energy per distance along $z$ can be derived from the second equation of Eq. 2,

$$
d(\gamma mc^2)/dz = -eA(z)E(r)\cos(\omega t - kz).
\tag{4}
$$

Here $t$ is the time that the electron arrives at $z$, but it depends on the history of the electron motion,

$$
t = t(z, t_0) = t_0 + \int_0^z \frac{ds}{u(s)},
\tag{5}
$$

which depends implicitly on the field and has to be solved by the first equation of Eq. 2.

Now we sum up the energy change for all electrons at the point $z$. Suppose the number of electrons per unit area entering the device at $t_0$ is $\frac{j_0}{e}dt_0$ where $j_0$ is the current density of the electrons prior to the entrance. The rate per unit area of these electrons arriving at $z$ is $\frac{j_0}{e}\frac{dt_0}{dt}|_z$. The current density at $z$ becomes $j_0 \frac{dt_0}{dt}|_z$; the same expression can be obtained by using the charge continuity. The description of bunching is contained in the term $\frac{dt_0}{dt}|_z = 1/\frac{dt(z,t_0)}{dt_0}$ that is determined by the electron motion.

The total radiation intensity change at $z$ is then given by

$$d(\gamma mc^2)/dz \frac{j_0}{e}\frac{dt_0}{dt}|_z$$
$$= -eA(z)E(r)\cos(\omega t - kz)\frac{j_0}{e}\frac{dt_0}{dt}|_z. \qquad (6)$$

Integrating over the beam cross section gives the power change per unit length for all electrons at $z$:

$$\begin{aligned}\frac{dw}{dz} &= \int_0^{r_b} 2\pi r dr \{-eA(z)E(r)\cos(\omega t - kz)\frac{j_0}{e}\frac{dt_0}{dt}|_z\} \\ &= -A(z)j_0 \frac{dt_0}{dt}|_z \cos(\omega t - kz)\int_0^{r_b} E(r)2\pi r dr, \end{aligned} \qquad (7)$$

where $w$ is the instantaneous radiation power, and $r_b$ is the electron beam diameter that should be not bigger than the vacuum tunnel radius $a$.

The root-mean-square (RMS) power ($W$) change is given by

$$\frac{dW}{dz} = -A(z)\int_0^{r_b} E(r)2\pi r dr \frac{1}{T}\int_{t-T/2}^{t+T/2} j_0 \frac{dt'_0}{dt'}|_z \cos(\omega t' - kz)dt' \qquad (8)$$

where $dt'$ becomes the integration variable and the integration is carried out over one cycle of the field oscillation $T = 2\pi/\omega$. It follows that $t' = t'(z, t'_0)$.

Now we calculate the radiation power change. Here instead of using the set of Maxwell equations, e.g. Eq. 1, to calculate the field properties from the electron motion [15], we calculate the EM field energy stored in the cavity:

$$w = \int_0^b E(\omega, k)H(\omega, k)2\pi r dr, \qquad (9)$$

where the transverse magnetic field is expressed by

$$H(\omega, k) = A(z)H(r)\cos(\omega t - kz), \qquad (10)$$

then the RMS radiation power change per unit length is given by

$$\frac{dW}{dz} = \frac{dA(z)}{dz}A(z)\int_0^b E(r)H(r)2\pi r dr. \qquad (11)$$

Equating the power change of the electrons (Eq. 8) and the power change of the fields (Eq. 11) based on energy conservation, we have the following equation:

$$-A(z)\int_0^{r_b} E(r)2\pi r dr \frac{1}{T}\int_{t-T/2}^{t+T/2} j_0 \frac{dt_0'}{dt'}|_z \cos(\omega t'-kz)dt' = \frac{dA(z)}{dz}A(z)\int_0^b E(r)H(r)2\pi r dr. \tag{12}$$

Rearranging Eq. 12 leads to the differential-integral equation:

$$\frac{dA(z)}{dz} = -\frac{\int_0^{r_b} E(r)2\pi r dr \frac{1}{T}\int_{t-T/2}^{t+T/2} j_0 \frac{dt_0'}{dt'}|_z \cos(\omega t'-kz)dt'}{\int_0^b E(r)H(r)2\pi r dr} \equiv -\frac{\int_0^{r_b} E(r)j_1(z)2\pi r dr}{\int_0^b E(r)H(r)2\pi r dr}. \tag{13}$$

We have thus derived the equation to calculate the function $A(z)$, the derivative of which is linked to the bunched current density:

$$\begin{aligned}j_1(z) &= \frac{1}{T}\int_{t-T/2}^{t+T/2} j_0 \frac{dt_0'}{dt'}|_z \cos(\omega t'-kz)dt' \\ &= \frac{1}{T}\int_{t_0-T/2}^{t_0+T/2} j_0 \cos(\omega t'-kz)dt_0'.\end{aligned} \tag{14}$$

The last step converts the integration over $dt'$ to $dt_0'$ but keeps the wave period $T$ the same as we assume the frequency of the field is unchanged during the interaction.

Equation 13 represents the key result obtained from this work. It states that the growth rate of the field amplitude over distance is determined by the ratio of the interaction power $\int_0^{r_b} E(r)j_1(z)2\pi r dr$ to the radiation power $\int_0^b E(r)H(r)2\pi r dr$. The latter is solely dependent on the cavity parameters, e.g., dielectric constant, geometry, etc. The interaction power, however, involves the current density $j_1(z)$. Coherent radiation of the same waveform $\cos(\omega t-kz)$ is generated by $j_1(z)$ and the power is proportional to $j_1(z)^2$. Assuming uniform spatial distribution of the current density, Eq. 13 can be further simplified by consolidating the cavity and initial beam characteristics into one term

$$\frac{dA(z)}{dz} = \frac{j_1(z)}{j_0}\Gamma, \tag{15}$$

where $\Gamma$ is a figure of merit of the device that characterizes the coupling strength,

$$\Gamma = \frac{\int_0^{r_b} E(r)j_0 2\pi r dr}{\int_0^b E(r)H(r)2\pi r dr}, \tag{16}$$

and $\frac{j_1(z)}{j_0} = \frac{1}{T}\int_{t_0-T/2}^{t_0+T/2} \cos(\omega t'-kz)dt_0'$ is the normalized bunched current density.

The dynamics of the device is now governed by Eqs. **??**, but requires the initial condition $A(0)$ to completely decide the device behavior. If the device is used as an amplifier, $A(0)$ is given by the external field at the entrance of the device. In the case of an oscillator, $A(0)$ is determined by the cavity properties, i.e. the reflectivity at the end or the cavity $Q$ value. The physical explanation is offered in the following. First, when the dc beam enters the device, it starts to generate spontaneous CR whose power is proportional to the number of electrons. Some of the generated radiation is reflected back to the cavity at the exit end and reflected again at the entrance end so that the electrons entering later start to interact with the "stored" field. The build-up of the field inside the cavity continues and oscillation is established if the following condition is satisfied

$$A(L)R_1R_2 \geq A(0), \qquad (17)$$

where $R_1$ and $R_2$ are the amplitude reflectivity at the two ends, and $L$ is the length of the cavity.

The gain, defined as $A(L)/A(0)$, obtained by recycling the spontaneous CR is too small to achieve oscillation, particularly for an open cavity structure like the one in the experiment [6]. However, the radiation is boosted once the self-bunching of the beam occurs—the beam-field interaction causes some electrons to speed up and some to slow down. The bunched electrons emit radiation coherently, and its power is proportional to the square of the number of electrons in the bunch; thus the radiation is considerably more intense than the spontaneous CR. Part of the generated radiation is reflected again into the cavity to bunch the newly arrived electrons. Therefore for oscillator, we need to add Eq. 17 to Eqs. ?? to determine the full device behavior. Notice that the generated field can not bunch the electron beam forever, and at certain point de-bunch starts to happen. This indicates that the field growth will reach a saturation point and even decay as evidence of the nonlinear behavior.

The same analysis leads to the explanation of fixing the oscillation frequency too. For most frequencies the normalized bunched current density $\frac{j_1(z)}{j_0}$ is either too small to contribute to the gain or simply negative. Physically the positive or negative bunched beam represents the electrons collectively generate or absorb the fields. To obtain monolithic gain of the field, the beam must be locked in phase to the radiation field and the only way to achieve that is to have the electrons travel at about the same velocity as the field phase velocity, $u = v_p = \omega/k$. This condition fixes the frequencies on the dispersion curve, which are known as the synchronous frequencies. This condition is similar to the "phase matching" condition in nonlinear optics upon which the highest gain occurs. The narrow-band radiation is predicted by the theory, and the actual oscillation frequencies will be shown to be slightly higher than the synchronous frequencies because the electrons have to travel a bit faster than the field for the CR effect to take place. It is evident that the oscillation frequencies can be tuned by adjusting the beam energy.

TABLE I: Device parameters

| Physical Quantity | Symbol | Cavity I |
|---|---|---|
| Inner radius | $a$ | 0.003175 m |
| Outer radius | $b$ | 0.00635 m |
| Refractive index | $n$ | 1.944 |
| Velocity | $u$ | $1.90 \times 10^8$ m/s |
| Current density | $j_0$ | $4.93 \times 10^5$ A/m$^2$ |

We now use the formalism developed here to evaluate the device used in the experiments [6]. The device parameters are listed in Table I. The current density is calculated for 10 A current with assumed beam size of $r_b = 0.8a$.

The electric field and transverse magnetic field $E(r)$ and $H(r)$ of the TM modes for the device are solved by the electron current-free Maxwell equations. The dispersion relation is also obtained as

$$\frac{I_1(Xa)}{I_0(Xa)X} = -\frac{\epsilon[Y_1(Ya)J_0(Yb) - J_1(Ya)Y_0(Yb)]}{\epsilon_0 Y[J_0(Ya)Y_0(Yb) - Y_0(Ya)J_0(Yb)]}, \quad (18)$$

where $J_n$, $I_n$ and $Y_n$ are the $n$th-order Bessel function of the first and second kind. $X = k^2 - (\frac{\omega}{c})^2$ and $Y = \frac{\epsilon}{\epsilon_0}(\frac{\omega}{c})^2 - k^2$ are the separation constants. The dispersion curves are shown in the inset of Fig. 3.

The dynamics of beam-field interaction inside the cavity is now calculated. First we check that the device can indeed build up from the noise (spontaneous CR) by assuming a very small $A(0)$; the device does display enough gain to satisfy the threshold of the oscillation condition (Eq. 17). When the device reaches the stable oscillation mode, the gain is pinched at the threshold oscillation condition, i.e. $A(L)/A(0) = 1/\sqrt{R_1 R_2}$, where $R_1 = R_2 = 0.125$ is estimated from the reflectivity calculated by a computer simulation, minus the scattering and absorption effects.

Figure 2 shows that the bunching of the current density $j_1(z)/j_0$ evolves over distance and the growth of the field amplitude $A(z)$, which largely follows the integration of the bunched current. The calculation is carried out over the distance longer than the device itself so that the nonlinear gain, e.g. saturation and decay, can be shown. The saturation occurs at the onset of the de-bunching of the beam, and the radiation power starts to decrease once the beam starts to bunch again, but out of phase with the field. This suggests that the cavity length needs to be optimized for maximum power output.

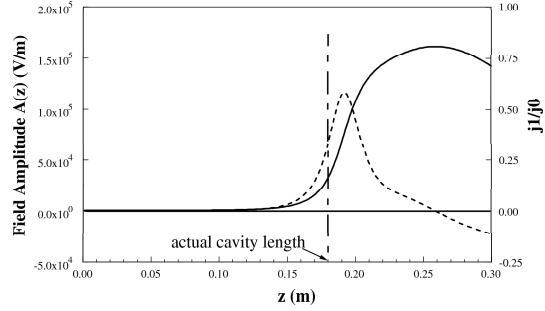

FIG. 2: Evolution of the field (solid line) and bunched current density (dashed line) over distance. The cavity length is 0.18 m. The frequency is selected to be 22.55 GHz and $A(0) = 500$.

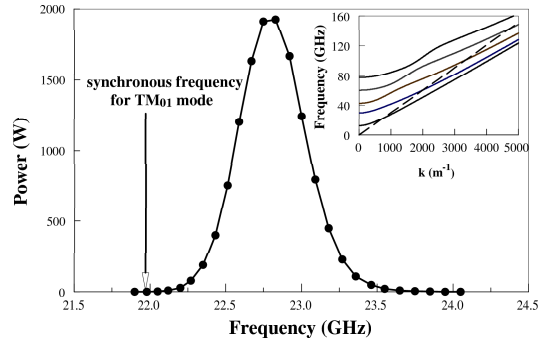

FIG. 3: The output radiation spectrum of the $TM_{01}$ mode at the exit of the device, L= 0.18m. The inset shows the dispersion curves (solid lines) and the synchronous frequencies as the interception points with the beam line (dotted).

The radiation spectrum is calculated by scanning the frequency around the synchronous point 21.90 GHz and the result is shown in Fig. 3. The bandwidth is measured to be about 500 MHz, which gives the averaged power to be about 1 kw, the same as the experimental result. The spectrum peaks at 22.8 GHz, about 750 MHz higher than the synchronous frequency. The frequency shift is observed by the experiment [6], but the experiment observes higher frequency shift than what the theory predicts.

The agreement with the experiment indicates that the theory could potentially be used to guide designing a working device at different frequencies. In general, the smaller the cavity, the higher the operation frequency. For example, if the cavity dimension is reduced to 2mm inner diameter and 2.4mm outer diameter, the first three synchronous frequencies are 296 GHz, 944 GHz and 1601 GHz, already in the THz range. To achieve the optimum device efficiency, the device geometry will have to be fine-tuned in conjunction with other factors such as the dielectric materials, thickness and the beam characteristics. The focus is to concentrate more electric field in the beam path as characterized by the $\Gamma$ factor. Materials of higher index of refraction leads to lower CR threshold, therefore lower electron beam energy can be used and higher energy conversion coefficient can be expected. Another interesting approach is to use periodic dielectric medium as discussed in [17], which will effectively lower the electron beam threshold to arbitrarily low values. Discussion on the detailed design issues will be presented elsewhere.

In short, we have developed the theory that describes the dynamics of the electron beam self-bunching and generation of coherent Cerenkov radiation from inside the cavity. We have shown that nonlinear gain demonstrates growth, saturation, and decay. We have derived the figure of merit $\Gamma$ (Eq. 16) of the device and predicted the output power spectrum that is consistent with the experimental result. The theory can be applied to other slow wave structures for both amplifiers and oscillators, however the emphasis of this paper is to show that the cavity CR device offers a viable approach for the pursuit of tunable THz radiation owing to its simplicity in structure and manufacturing, and the ability to be integrated with the many nano-enhanced electron field emitters [18].

The author thanks stimulating discussion with P. D. Coleman and J.T. Verdeyen.

[1] R. Khler, A. Tredicucci, F. Beltram, H. E. Beere, E. H. Linfield, A. G. Davies, and D. A. Ritchie, Optics Letters **28**, 810 (2003).

[2] L. Mallet, C.R. Acad. Sci. (Paris) **183**, 274 (1926).

[3] P. A. Cerenkov, Compt. Rend. **2**, 451 (1934).

[4] V. L. Ginzburg, J. Phys. (U.R.S.S) **2**, 441 (1940).

[5] P. D. Coleman and C. Enderby, J. Appl. Phys. **31**, 1699 (1960).

[6] K. L. Felsh, K. O. Busby, R. W. Layman, D. Kapilow, and J. E. Walsh, Appl. Phys. Lett. **38**, 601 (1981).

[7] J. T. Verdeyen, *Laser Electronics* (Prentice Hall, New Jersey, 1995).

[8] C. A. Brau, *Free-Electron Lasers* (Academic Press, 1990).

[9] J. R. Pierce, *Travelling-wave Tubes* (D. Van Nostrand, 1950).

[10] L. J. Chu and J. D. Jackson, Proc. I.R.E. **36**, 853 (1948).

[11] B. M. Bolotvskii and U. F. Nauk, Sov. Phys. Usp. **4**, 781 (1962).

[12] E. Garate, R. Cook, P. Heim, R. W. Layman, and J. E. Walsh, **58** (1985).

[13] J. E. Walsh and J. B. Murphy, IEEE J. Quantum Electron **8**, 1259 (1982).

[14] B. Johnson and J. E. Walsh, Phys. Rev. A **33**, 3199 (1986).

[15] L. Schachter, Phys. Rev. A **43**, 3785 (1991).

[16] L. Schachter and J. A. Nation, Phys. Rev. A **45**, 8820 (1992).

[17] P. D. Coleman, M. Lerttamrab, and J. Gao, Phys. Rev. E **66**, 066501 (2002).

[18] D. S. Y. Hsu and J. Shaw, Appl. Phys. Lett. **80** (2002).